# Towards a new crown indicator: An empirical analysis

Ludo Waltman, Nees Jan van Eck, Thed N. van Leeuwen,
Martijn S. Visser, and Anthony F.J. van Raan

Centre for Science and Technology Studies, Leiden University, The Netherlands
{waltmanlr, ecknjpvan, leeuwen, visser, vanraan}@cwts.leidenuniv.nl

We present an empirical comparison between two normalization mechanisms for citation-based indicators of research performance. These mechanisms aim to normalize citation counts for the field and the year in which a publication was published. One mechanism is applied in the current so-called crown indicator of our institute. The other mechanism is applied in the new crown indicator that our institute is planning to adopt. We find that at high aggregation levels, such as at the level of large research institutions or at the level of countries, the differences between the two mechanisms are very small. At lower aggregation levels, such as at the level of research groups or at the level of journals, the differences between the two mechanisms are somewhat larger. We pay special attention to the way in which recent publications are handled. These publications typically have very low citation counts and should therefore be handled with special care.

## 1. Introduction

It is well known that the average number of citations per publication varies significantly across scientific fields. Of course, the average number of citations per publication also varies across publications of different ages. That is, older publications on average have more citations than newer ones. Due to these effects, citation counts of publications published in different fields or in different years cannot be directly compared with each other.

It is generally agreed that in citation-based research performance evaluations one needs to control for the field and the year in which a publication was published. In performance evaluation studies, our institute, the Centre for Science and Technology Studies (CWTS) of Leiden University, uses a standard set of bibliometric indicators (Van Raan, 2005). Our best-known indicator, which we usually refer to as the crown indicator, relies on a normalization mechanism that aims to correct for the field and the year in which a publication was published. An indicator similar to the crown indicator is used by the Centre for R&D Monitoring (ECOOM) in Leuven, Belgium. ECOOM calls its indicator the normalized mean citation rate (e.g., Glänzel, Thijs, Schubert, & Debackere, 2009).

The normalization mechanism of the crown indicator basically works as follows. Given a set of publications, we count for each publication the number of citations it has received. We also determine for each publication its expected number of citations. The expected number of citations of a publication equals the average number of citations of all publications of the same document type (i.e., article, letter, or review) published in the same field and in the same year. To obtain the crown indicator, we divide the sum of the actual number of citations of all publications by the sum of the expected number of citations of all publications.

As an alternative to the above normalization mechanism, one could take the following approach. One first calculates for each publication the ratio of its actual number of citations and its expected number of citations, and one then takes the average of the ratios that one has obtained. An indicator that uses this alternative normalization mechanism was introduced by Lundberg (2007). He called his indicator the item-oriented field-normalized citation score average. More recently, Opthof and Leydesdorff (2010) argued in favor of the alternative normalization mechanism.[1] Indicators that rely on the alternative mechanism are being used by various institutes, among which Karolinska Institute in Sweden (Rehn & Kronman, 2008), Science-Metrix in the US and Canada (e.g., Campbell, Archambault, & Côté, 2008, p. 12), the SCImago research group in Spain (SCImago Research Group, 2009), and Wageningen University in the Netherlands (Van Veller, Gerritsma, Van der Togt, Leon, & Van Zeist, 2009). Sandström also employed the alternative mechanism in various bibliometric studies (e.g., Sandström, 2009, p. 33–34).

In a recent paper (Waltman, Van Eck, Van Leeuwen, Visser, & Van Raan, in press), we have presented a theoretical comparison between the normalization mechanism of the crown indicator and the alternative normalization mechanism discussed by Lundberg (2007) and others. The main conclusion that we have reached is that, at least for the purpose of correcting for the field in which a publication was published, the alternative mechanism has more satisfactory properties than the mechanism of the crown indicator. In particular, the alternative mechanism weighs all publications equally while the mechanism of the crown indicator gives more weight to publications from fields with a large expected number of citations. The alternative mechanism also has a so-called consistency property. Basically, this property ensures that the ranking of two units relative to each other does not change when both units make the same progress in terms of publications and citations. The normalization mechanism of the crown indicator does not have this important property.

At CWTS, we are currently moving towards a new crown indicator, in which we use the alternative normalization mechanism. In this paper, we explore the consequences of this change. We perform an empirical comparison between on the one hand the normalization mechanism of our current crown indicator and on the other hand the alternative normalization mechanism that we are going to use in our new crown indicator. Our focus in this paper is on the issue of correcting for the field and the year in which a publication was published. We do not consider the issue of correcting for a publication's document type. We study four aggregation levels at which bibliometric indicators can be calculated, namely the level of research groups, the level of research institutions, the level of countries, and the level of journals. We pay special attention to the way in which recent publications are handled when the alternative normalization mechanism is used.

## 2. Definitions of indicators

In this section, we formally define the CPP/FCSm indicator and the MNCS indicator. The CPP/FCSm indicator, where CPP and FCSm are acronyms for, respectively, citations per publication and mean field citation score, has been used as the crown indicator of CWTS for more than a decade. The MNCS indicator, where MNCS is an acronym for mean normalized citation score, is the new crown indicator that CWTS is going to adopt.

Consider a set of $n$ publications, denoted by 1, …, $n$. Let $c_i$ denote the number of citations of publication $i$, and let $e_i$ denote the expected number of citations of publication $i$ given the field and the year in which publication $i$ was published. In other words, $e_i$ equals the average number of citations of all publications published in the same field and in the same year as publication $i$. The field in which a publication was published can be defined in many different ways. At CWTS, we normally define fields based on subject categories in the Web of Science database. The CPP/FCSm indicator is defined as

$$\text{CPP/FCSm} = \frac{\sum_{i=1}^{n} c_i / n}{\sum_{i=1}^{n} e_i / n} = \frac{\sum_{i=1}^{n} c_i}{\sum_{i=1}^{n} e_i}. \quad (1)$$

[1] See also our reply to Opthof and Leydesdorff (Van Raan, Van Leeuwen, Visser, Van Eck, & Waltman, 2010) and some other contributions to the discussion (Bornmann, 2010; Moed, 2010; Spaan, 2010).

The CPP/FCSm indicator was introduced by De Bruin, Kint, Luwel, and Moed (1993) and Moed, De Bruin, and Van Leeuwen (1995). A similar indicator, the normalized mean citation rate, was introduced somewhat earlier by Braun and Glänzel (1990).[2] The normalization mechanism of the CPP/FCSm indicator goes back to Schubert and Braun (1986) and Vinkler (1986). For a discussion of the conceptual foundation of the CPP/FCSm indicator, we refer to Moed (2010).

We now turn to the MNCS indicator (Waltman et al., in press). This indicator is defined as

$$\text{MNCS} = \frac{1}{n} \sum_{i=1}^{n} \frac{c_i}{e_i}. \quad (2)$$

The MNCS indicator is similar to the item-oriented field-normalized citation score average indicator introduced by Lundberg (2007). The normalization mechanism of the MNCS indicator is also applied in the relative paper citation rate indicator discussed by Vinkler (1996). Comparing (1) and (2), it can be seen that the CPP/FCSm indicator normalizes by calculating a ratio of averages while the MNCS indicator normalizes by calculating an average of ratios.

## 3. How to handle recent publications?

Recent publications usually do not have much effect in the calculation of the CPP/FCSm indicator. These publications tend to have no more than a few citations, and they also tend to have a low expected number of citations. As a consequence, both in the numerator and in the denominator of the CPP/FCSm indicator, the effect of recent publications is typically quite small.[3] This is different in the case of the MNCS indicator. Unlike the CPP/FCSm indicator, the MNCS indicator weighs all publications equally. Because of this, recent publications have an equally strong effect in the calculation of the MNCS indicator as older publications.

Weighing all publications equally seems very natural and has theoretical advantages (Waltman et al., in press). However, it also has a disadvantage. Recent publications have not had much time to earn citations, and their current number of citations therefore need not be a very accurate indicator of their long-run impact. We now present some empirical data to illustrate this issue.

---

[2] The difference between the normalized mean citation rate indicator and the CPP/FCSm indicator is that the former indicator only normalizes for the field and the year in which a publication was published while the latter indicator also normalizes for a publication's document type. In this paper, we do not consider the issue of normalizing for a publication's document type. For our present purpose, the difference between the two indicators is therefore not important.

[3] An alternative explanation is as follows. It can be shown mathematically that the CPP/FCSm indicator weighs publications proportionally to their expected number of citations (Waltman et al., in press, Section 2). Recent publications tend to have a low expected number of citations, and their effect in the calculation of the CPP/FCSm indicator therefore tends to be small.

Our analysis is based on the Web of Science database. We selected seven subject categories in this database. We interpret these subject categories as scientific fields. The selected subject categories are listed in the first column of Table 1. For each of the selected subject categories, we identified all publications of the document types article and review published in 1999 in journals belonging to the subject category. For each of the identified publications, we counted the number of times the publication had been cited by the end of each year between 1999 and 2008. Author self-citations are not included in our citation counts. For each subject category, the number of identified publications is listed in the second column of Table 1. Average citation counts of the identified publications are reported in the remaining columns of the table.

Table 1. Average citation counts of publications published in 1999 in seven subject categories.

| | No of pub. | Average number of citations per publication by the end of 1999 | 2000 | 2001 | 2002 | 2003 | 2004 | 2005 | 2006 | 2007 | 2008 |
|---|---|---|---|---|---|---|---|---|---|---|---|
| Biochemistry & molecular biology | 45,721 | 0.5 | 3.4 | 7.3 | 11.0 | 14.5 | 17.9 | 20.9 | 23.8 | 26.4 | 28.9 |
| Cardiac & cardiovascular systems | 11,332 | 0.3 | 2.0 | 4.7 | 7.4 | 10.0 | 12.6 | 14.9 | 17.0 | 19.1 | 20.9 |
| Chemistry, analytical | 13,887 | 0.1 | 1.1 | 2.5 | 4.0 | 5.5 | 7.0 | 8.5 | 10.0 | 11.4 | 12.7 |
| Economics | 7,346 | 0.1 | 0.5 | 1.2 | 2.0 | 3.0 | 4.1 | 5.3 | 6.5 | 7.9 | 9.4 |
| Mathematics | 12,450 | 0.0 | 0.2 | 0.5 | 0.8 | 1.2 | 1.6 | 2.1 | 2.5 | 2.9 | 3.4 |
| Physics, applied | 24,675 | 0.1 | 0.7 | 1.7 | 2.8 | 3.9 | 4.9 | 6.0 | 7.0 | 8.0 | 8.8 |
| Surgery | 22,230 | 0.1 | 0.9 | 2.4 | 3.9 | 5.4 | 6.9 | 8.3 | 9.6 | 11.0 | 12.3 |

The citation counts in Table 1 show large differences among fields. Biochemistry and molecular biology has the highest citation counts, and mathematics has the lowest. The difference is roughly one order of magnitude. This difference clearly indicates the importance of correcting for the field in which a publication was published. It can further be seen in Table 1 that during the first ten years after a publication was published citation counts on average increase approximately linearly with time.

As shown in the third column of Table 1, publications receive almost no citations in the year in which they were published. This is not surprising. Citing publications need to be written, reviewed, revised, and copyedited, which even under the most favorable conditions takes at least several months. In addition, some journals have a substantial backlog of manuscripts waiting to be published. This also delays the citation process. For these reasons, it is unlikely that publications receive more than a few citations in the year in which they were published.[4] This is especially true for publications published towards the end of the year. Notice in Table 1 that in some fields, in particular in mathematics, publications are unlikely to be cited not only in the year in which they were published but also in the next year.

[4] However, as we will see later on in this paper, there are exceptional publications that receive lots of citations already in the year in which they were published.

How well does the number of citations of a publication one or two years after the publication appeared predict the number of citations of the publication in the medium or long run, say, after five or ten years? In Table 2, we report for any two years $y_1$ and $y_2$, with $y_1$ and $y_2$ between 1999 and 2008, the Pearson correlation between the number of citations a publication has received by the end of year $y_1$ and the number of citations a publication has received by the end of year $y_2$. The correlations in the upper right part of the table were calculated for publications published in 1999 in biochemistry and molecular biology journals. The correlations in the lower left part of the table were calculated for publications published in 1999 in mathematics journals.

Table 2. Pearson correlations between the number of citations a publication has received by the end of one year and the number of citations a publication has received by the end of another year. The upper right part and the lower left part of the table relate to publications published in 1999 in, respectively, biochemistry and molecular biology journals and mathematics journals.

| | 1999 | 2000 | 2001 | 2002 | 2003 | 2004 | 2005 | 2006 | 2007 | 2008 |
|---|---|---|---|---|---|---|---|---|---|---|
| 1999 | | 0.83 | 0.74 | 0.68 | 0.65 | 0.62 | 0.60 | 0.58 | 0.56 | 0.55 |
| 2000 | 0.56 | | 0.96 | 0.93 | 0.90 | 0.87 | 0.85 | 0.83 | 0.81 | 0.79 |
| 2001 | 0.43 | 0.82 | | 0.99 | 0.97 | 0.95 | 0.93 | 0.92 | 0.90 | 0.88 |
| 2002 | 0.37 | 0.74 | 0.92 | | 0.99 | 0.98 | 0.97 | 0.96 | 0.94 | 0.93 |
| 2003 | 0.33 | 0.70 | 0.87 | 0.96 | | 1.00 | 0.99 | 0.98 | 0.97 | 0.95 |
| 2004 | 0.31 | 0.67 | 0.83 | 0.92 | 0.97 | | 1.00 | 0.99 | 0.98 | 0.97 |
| 2005 | 0.29 | 0.64 | 0.80 | 0.89 | 0.95 | 0.98 | | 1.00 | 0.99 | 0.99 |
| 2006 | 0.28 | 0.62 | 0.78 | 0.87 | 0.93 | 0.97 | 0.99 | | 1.00 | 0.99 |
| 2007 | 0.26 | 0.60 | 0.75 | 0.85 | 0.91 | 0.95 | 0.98 | 0.99 | | 1.00 |
| 2008 | 0.25 | 0.59 | 0.74 | 0.83 | 0.89 | 0.93 | 0.96 | 0.98 | 0.99 | |

As can be seen in Table 2, correlations between short-run citation counts and long-run citation counts can be quite weak. In the case of mathematics publications published in 1999, the correlation between the number of citations received by the end of 1999 and the number of citations received by the end of 2008 equals just 0.25. The correlation between the number of citations received by the end of 2000 and the number of citations received by the end of 2008 equals 0.59, which is still only a very moderate correlation. Of the seven subject categories that we have selected, biochemistry and molecular biology has the strongest correlations between short-run citation counts and long-run citation counts. This is to be expected, since biochemistry and molecular biology also has the highest citation counts. However, even in the case of biochemistry and molecular biology publications, the correlation between the number of citations received by the end of 1999 and the number of citations received by the end of 2008 is rather moderate, with a value of just 0.55.

Based on Tables 1 and 2, we conclude that in the calculation of the MNCS indicator recent publications need special attention. These publications have low citation counts (Table 1), and because of this their long-run impact cannot be predicted very well (Table 2). This is not a big problem in the case of the CPP/FCSm indicator, since this indicator gives less weight to recent publications than to older ones. The MNCS indicator, however, weighs all publications equally, and recent publications may then introduce a quite significant amount of noise in the indicator. Especially when the MNCS indicator is calculated at lower aggregation levels (e.g., at the level of research groups or individual researchers), where only a limited number of publications are available, this can be a serious problem. To alleviate this problem, one may consider leaving out the most recent publications in the calculation of the MNCS indicator. For example, all publications that have had less than one year to earn citations could be left out. In this way, one loses some relevant information, but one also gets rid of a lot of noise.

## 4. Empirical comparison

In this section, we present an empirical comparison between the CPP/FCSm indicator and the MNCS indicator. We distinguish between two variants of the MNCS indicator. In one variant, referred to as the MNCS1 indicator, all publications are taken into consideration. In the other variant, referred to as the MNCS2 indicator, publications that have had less than one year to earn citations are left out.

We study four aggregation levels at which bibliometric indicators can be calculated, namely the level of research groups, the level of research institutions, the level of countries, and the level of journals. We do not consider the level of individual researchers. An analysis at this level can be found elsewhere (Van Raan et al., 2010). We use the following four data sets:

- *Research groups*. Chemistry and chemical engineering research groups in the Netherlands. This data set has been employed in a performance evaluation study for the Association of Universities in the Netherlands (VSNU, 2002).
- *Research institutions*. The 365 universities with the largest number of publications in the Web of Science database.
- *Countries*. The 58 countries with the largest number of publications in the Web of Science database.
- *Journals*. All journals in the Web of Science database except arts and humanities journals.

The main characteristics of the data sets are listed in Table 3.

Table 3. Characteristics of the data sets used to compare the CPP/FCSm indicator and the MNCS indicator.

| | Research groups | Research institutions | Countries | Journals |
|---|---|---|---|---|
| *N* | 158 | 365 | 58 | 8,423 |
| Time period | 1991–2000 | 2001–2008 | 2001–2008 | 2005–2008 |
| Average no. of pub. | 131 | 15,069 | 154,512 | 475 |
| Median no. of pub. | 103 | 12,409 | 47,506 | 233 |
| St. dev. no. of pub. | 103 | 9,149 | 325,787 | 1,027 |

The comparison between the CPP/FCSm indicator and the MNCS indicator was performed as follows. For each research group, research institution, country, or journal, we retrieved from the Web of Science database all publications of the document types article, note, and review published in the relevant time period specified in Table 3. Publications in the arts and humanities were left out of the analysis. This was done because these publications tend to have very low citation counts, which makes the use of citation-based performance indicators problematic. We counted citations until the end of the relevant time period. Author self-citations were ignored. In the calculation of the indicators, we normalized for the field and the year in which a publication was published. We did not normalize for a publication's document type. Fields were defined by Web of Science subject categories. As mentioned earlier, in the MNCS2 indicator, publications that have had less than one year to earn citations are left out. In the other two indicators, all publications are taken into consideration.

For each of the four data sets that we use, Pearson and Spearman correlations between the CPP/FCSm indicator, the MNCS1 indicator, and the MNCS2 indicator are reported in Table 4. The Pearson correlation measures to what degree two indicators are linearly related. The Spearman correlation, on the other hand, measures to what degree two indicators are monotonically related (i.e., to what degree two indicators yield the same ranking of items). Scatter plots of the relations between the indicators are shown in Figures 1 to 10. Items with no more than 50 publications (excluding publications that have had less than one year to earn citations) are indicated by red squares in these figures. Items with more than 50 publications are indicated by blue circles. In each figure, a 45-degree line through the origin has been drawn. The closer items are located to this line, the stronger the relation between two indicators.

Table 4. Pearson and Spearman correlations between the CPP/FCSm indicator, the MNCS1 indicator, and the MNCS2 indicator.

| | Research groups | Research institutions | Countries | Journals |
|---|---|---|---|---|
| CPP/FCSm vs MNCS1 (Pearson) | 0.85 | 0.98 | 0.99 | 0.94 |
| CPP/FCSm vs MNCS1 (Spearman) | 0.89 | 0.98 | 0.99 | 0.95 |
| CPP/FCSm vs MNCS2 (Pearson) | 0.91 | 0.99 | 0.99 | 0.96 |
| CPP/FCSm vs MNCS2 (Spearman) | 0.95 | 0.99 | 0.99 | 0.98 |
| MNCS1 vs MNCS2 (Pearson) | 0.95 | 0.99 | 1.00 | 0.91 |
| MNCS1 vs MNCS2 (Spearman) | 0.95 | 0.99 | 1.00 | 0.96 |

We first consider the research groups data set. For this data set, we observe a moderately strong relation between the CPP/FCSm indicator and the MNCS1 indicator (see Figure 1). For most research groups, the difference between the CPP/FCSm score and the MNCS1 score is not very large. However, there are a number of research groups for which the MNCS1 score is much higher or much lower than the CPP/FCSm score. The relation between the CPP/FCSm indicator and the MNCS2 indicator is considerably stronger (see Figure 2). There are only a small number of research groups for which the CPP/FCSm score and the MNCS2 score really differ significantly from each other. The three research groups for which the difference is largest have been marked with the letters A, B, and C in Figure 2. Let us consider these research groups in more detail. Research group A has only 15 publications. For each of these publications, we report in Table 5 the publication year, the number of citations, the expected number of citations, and the normalized citation score. The normalized citation score of a publication is defined as the ratio of the actual and the expected number of citations of the publication. Why is the CPP/FCSm score of research group A so much lower than the MNCS2 score of this research group? As can be seen in Table 5, the three publications of research group A with the highest normalized citation score were all published in 1999, which is second-last year of the analysis. These publications have a large effect on the MNCS2 score of research group A.[5] Their effect on the CPP/FCSm score of research group A is much smaller. This is because, as discussed earlier, recent publications have less weight in the CPP/FCSm indicator than in the MNCS2 indicator. This explains why the CPP/FCSm score of research group A is much lower than the MNCS2 score. Research groups B and C have more publications than research group A (respectively 42 and 165), but the explanation for the difference between the CPP/FCSm score and the MNCS2 score is similar. Like research group A, research group B has a number of recent publications with a high normalized citation score. Because of this, the MNCS2 score of research group B is much higher than the CPP/FCSm score. Research group C has two very highly cited publications in 1991, the first year of the analysis. These publications have more weight in the CPP/FCSm indicator than in the MNCS2 indicator, which explains the difference between the CPP/FCSm score and the MNCS2 score of research group C.

[5] Notice in Table 5 that the publication with the highest normalized citation score has just five citations. The high normalized citation score of this publication is due to the small expected number of citations of the publication. This illustrates that in the calculation of the MNCS2 indicator a recent publication with a relatively small number of citations can already have a quite large effect.

Table 5. Publication year, number of citations, expected number of citations, and normalized citation score of the publications of research group A.

| Pub. year | No of cit. | Exp. no of cit. | Norm. cit. score |
|---|---|---|---|
| 1994 | 6 | 6.97 | 0.86 |
| 1994 | 3 | 6.97 | 0.43 |
| 1995 | 0 | 7.39 | 0.00 |
| 1995 | 2 | 2.54 | 0.79 |
| 1995 | 5 | 7.39 | 0.68 |
| 1997 | 21 | 3.57 | 5.89 |
| 1997 | 1 | 4.42 | 0.23 |
| 1998 | 6 | 2.48 | 2.42 |
| 1998 | 6 | 2.48 | 2.42 |
| 1998 | 3 | 2.17 | 1.38 |
| 1999 | 16 | 1.52 | 10.55 |
| 1999 | 13 | 1.52 | 8.57 |
| 1999 | 5 | 0.45 | 11.03 |
| 1999 | 1 | 1.09 | 0.91 |
| 2000 | 0 | 0.21 | 0.00 |

We now turn to the research institutions data set. For this data set, we observe a very strong relation between on the one hand the CPP/FCSm indicator and on the other hand the MNCS1 indicator and the MNCS2 indicator (see Figures 3 and 4). The relation is approximately equally strong for both MNCS variants. As can be seen in Figure 3, there is one university for which the MNCS1 score (1.66) is much higher than the CPP/FCSm score (1.06). It turns out that in 2008 this university, the University of Göttingen, published an article that by the end of 2008 had already been cited 3489 times.[6] Since this is a very recent article, it has much more weight in the MNCS1 indicator than in the CPP/FCSm indicator. This explains the very different CPP/FCSm and MNCS1 scores of the university. Notice that in the MNCS2 indicator articles published in 2008 are not taken into consideration. Because of this, there is no substantial difference between the CPP/FCSm score (1.06) and the MNCS2 score (1.10) of the university.

The results obtained for the countries data set are similar to those obtained for the research institutions data set. We again observe a very strong relation between the CPP/FCSm indicator and the two MNCS variants (see Figures 5 and 6), and again the relation is approximately equally strong for both MNCS variants. A striking observation is that there are almost no countries for which the MNCS1 and MNCS2 scores are lower than the CPP/FCSm score. We currently do not have an explanation for this observation. In Table 6, we list the ten highest-ranked countries according to each of the three indicators that we study. As can be seen, the three indicators yield very similar results.

[6] The extremely large number of citations of this recently published article was also discussed by Dimitrov, Kaveri, and Bayry (2010), who pointed out the enormous effect of this single article on the impact factor of *Acta Crystallographica Section A*, the journal in which the article was published.

Table 6. The ten highest-ranked countries according to the CPP/FCSm indicator, the MNCS1 indicator, and the MNCS2 indicator.

| Rank | Country | CPP/FCSm | Country | MNCS1 | Country | MNCS2 |
|---|---|---|---|---|---|---|
| 1 | Switzerland | 1.43 | Switzerland | 1.47 | Switzerland | 1.45 |
| 2 | USA | 1.38 | USA | 1.39 | USA | 1.38 |
| 3 | Netherlands | 1.34 | Denmark | 1.37 | Netherlands | 1.36 |
| 4 | Denmark | 1.31 | Netherlands | 1.37 | Denmark | 1.34 |
| 5 | UK | 1.27 | UK | 1.29 | UK | 1.27 |
| 6 | Ireland | 1.23 | Sweden | 1.24 | Sweden | 1.23 |
| 7 | Canada | 1.22 | Belgium | 1.22 | Belgium | 1.21 |
| 8 | Belgium | 1.21 | Canada | 1.21 | Canada | 1.21 |
| 9 | Sweden | 1.20 | Ireland | 1.20 | Ireland | 1.21 |
| 10 | Norway | 1.18 | Norway | 1.19 | Norway | 1.20 |

Finally, we turn to the journals data set. For a large majority of the journals, we observe a strong relation between the CPP/FCSm indicator and the MNCS1 indicator (see Figures 7 and 8).[7] However, there are also a substantial number of journals for which the MNCS1 score is much higher or much lower than the CPP/FCSm score. Comparing the CPP/FCSm indicator with the MNCS2 indicator, we observe much less journals with largely different scores (see Figures 9 and 10). Hence, the CPP/FCSm indicator has a considerably stronger relation with the MNCS2 indicator than with the MNCS1 indicator. This is similar to what we found for the research groups data set. Notice that even when CPP/FCSm scores are compared with MNCS2 scores, there are a number of journals for which rather large differences can be observed. However, given that overall we have more than 8000 journals, these journals constitute a small minority of exceptional cases.[8]

## 5. Conclusions

We have presented an empirical comparison between two normalization mechanisms for citation-based indicators of research performance. One normalization mechanism is implemented in the CPP/FCSm indicator, which is the current crown indicator of CWTS. The other normalization mechanism is implemented in the MNCS indicator, which is the new crown indicator that CWTS is going to adopt. Our empirical results indicate that at high aggregation levels, such as at the level of large research institutions or at the level of countries, the differences between the CPP/FCSm indicator and the MNCS indicator are very small. At lower aggregation levels, such as at the level of research groups or at the level of journals, the differences between the two indicators are somewhat larger.

[7] In the case of journals, the CPP/FCSm indicator is also referred to as the JFIS indicator (e.g., Van Leeuwen & Moed, 2002).

[8] Comparing Figures 7 and 9, it can be seen that the journal with the highest CPP/FCSm score (17.68) has extremely different MNCS1 and MNCS2 scores (respectively 32.28 and 2.14). The MNCS1 score of the journal is much higher than the CPP/FCSm score, while the MNCS2 score is much lower. It turns out that in 2008 the journal, *Acta Crystallographica Section A*, published an article that by the end of 2008 had already been cited 3489 times. This is the same article mentioned earlier for the University of Göttingen. This article has much more weight in the MNCS1 indicator than in the CPP/FCSm indicator. In the MNCS2 indicator, the article is not taken into consideration at all. This explains the extremely different CPP/FCSm, MNCS1, and MNCS2 scores of the journal.

We have also pointed out that recent publications need special attention in the calculation of the MNCS indicator. These publications have low citation counts, and because of this their long-run impact cannot be predicted very well. Since the MNCS indicator gives the same weight to recent publications as to older ones, recent publications may introduce a significant amount of noise in this indicator. To alleviate this problem, one may consider leaving out the most recent publications in the calculation of the indicator. In our empirical analysis, we have examined the effect of leaving out publications that have had less than one year to earn citations. At lower aggregation levels, the effect turns out to be quite substantial. In particular, leaving out the most recent publications in the calculation of the MNCS indicator turns out to lead to a stronger relation between the CPP/FCSm indicator and the MNCS indicator. This suggests that differences between the CPP/FCSm indicator and the MNCS indicator may be partly due to noise introduced in the MNCS indicator by recent publications.

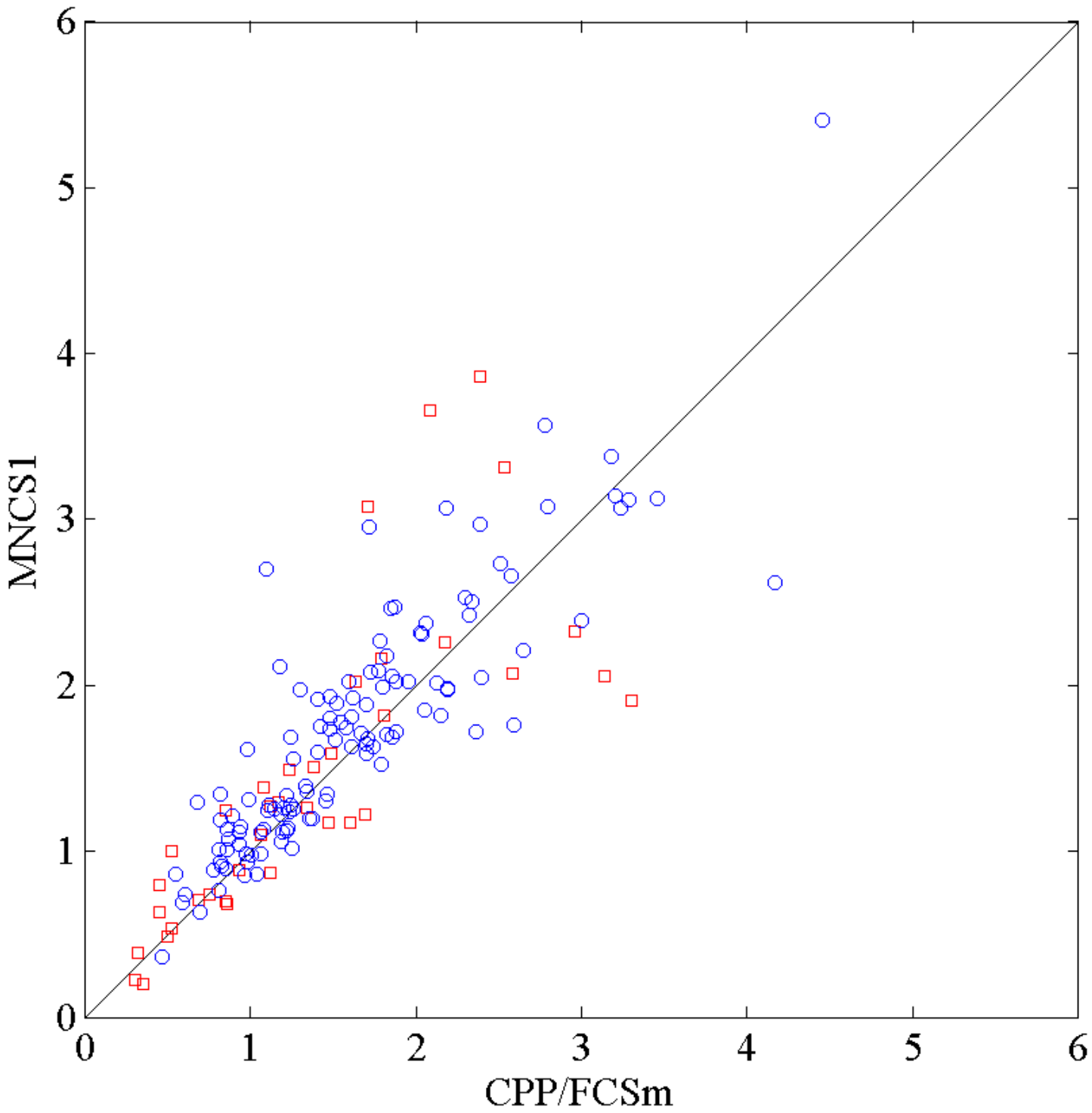

Figure 1. Relation between the CPP/FCSm indicator and the MNCS1 indicator for the research groups data set.

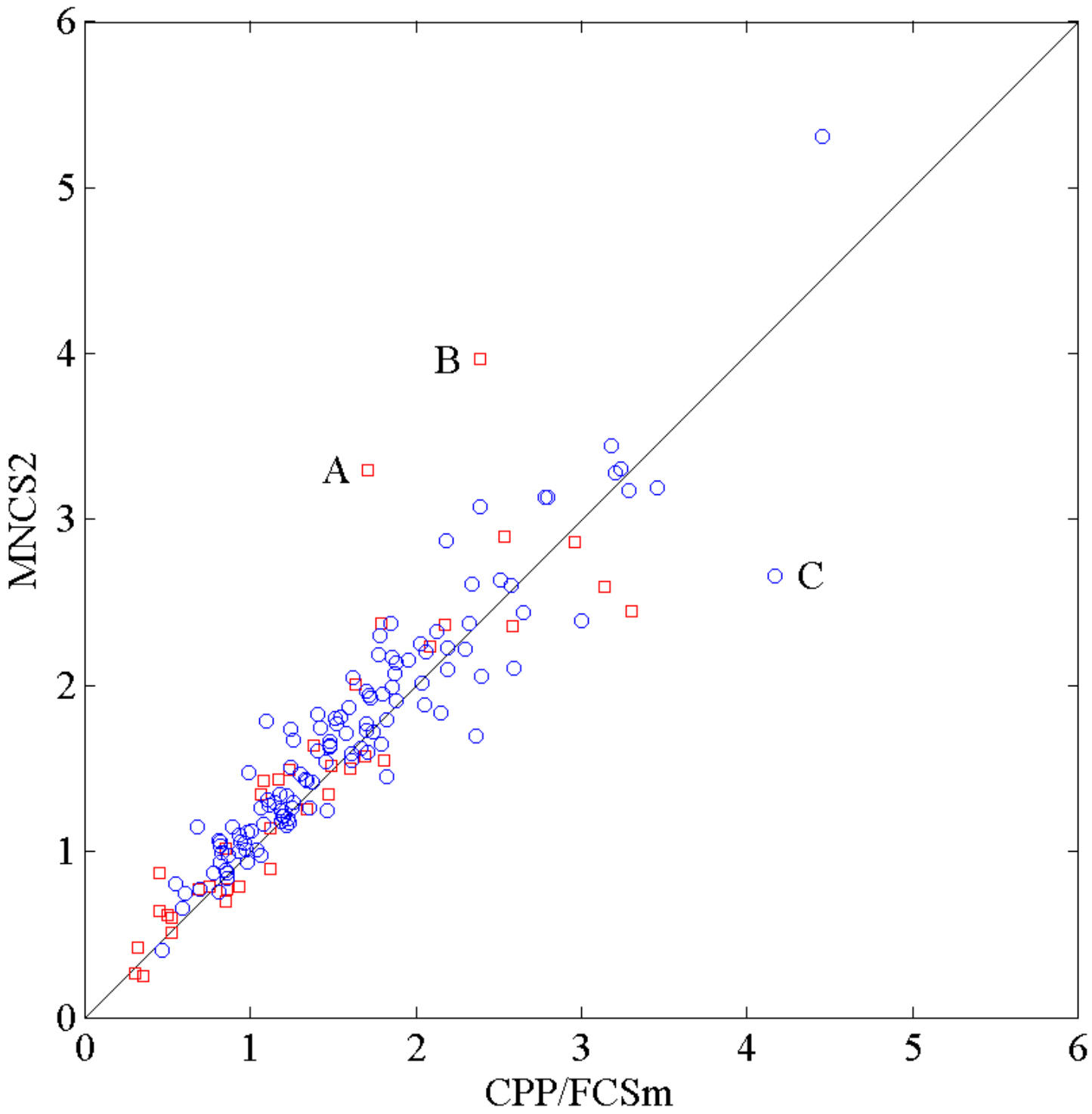

Figure 2. Relation between the CPP/FCSm indicator and the MNCS2 indicator for the research groups data set.

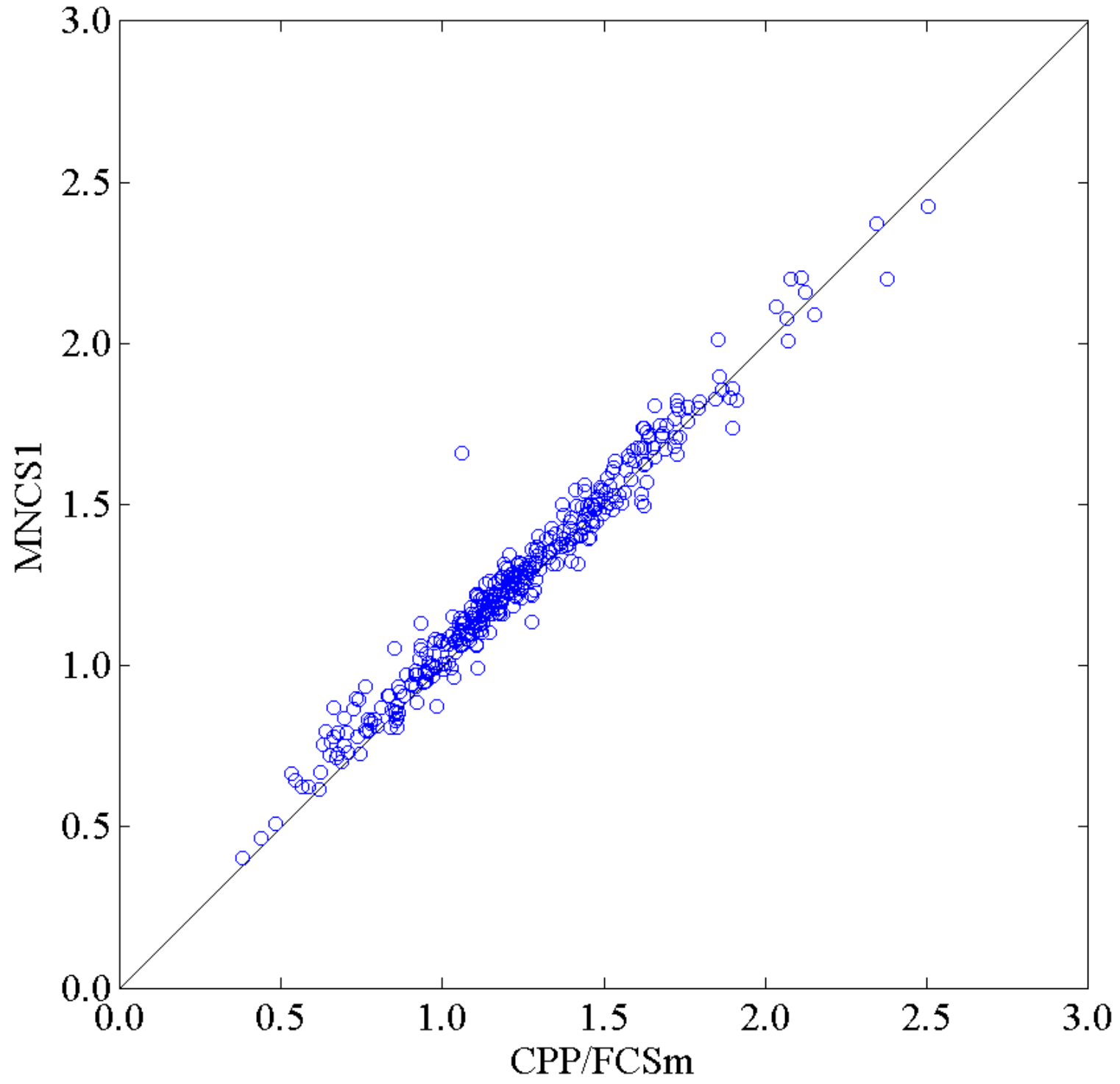

Figure 3. Relation between the CPP/FCSm indicator and the MNCS1 indicator for the research institutions data set.

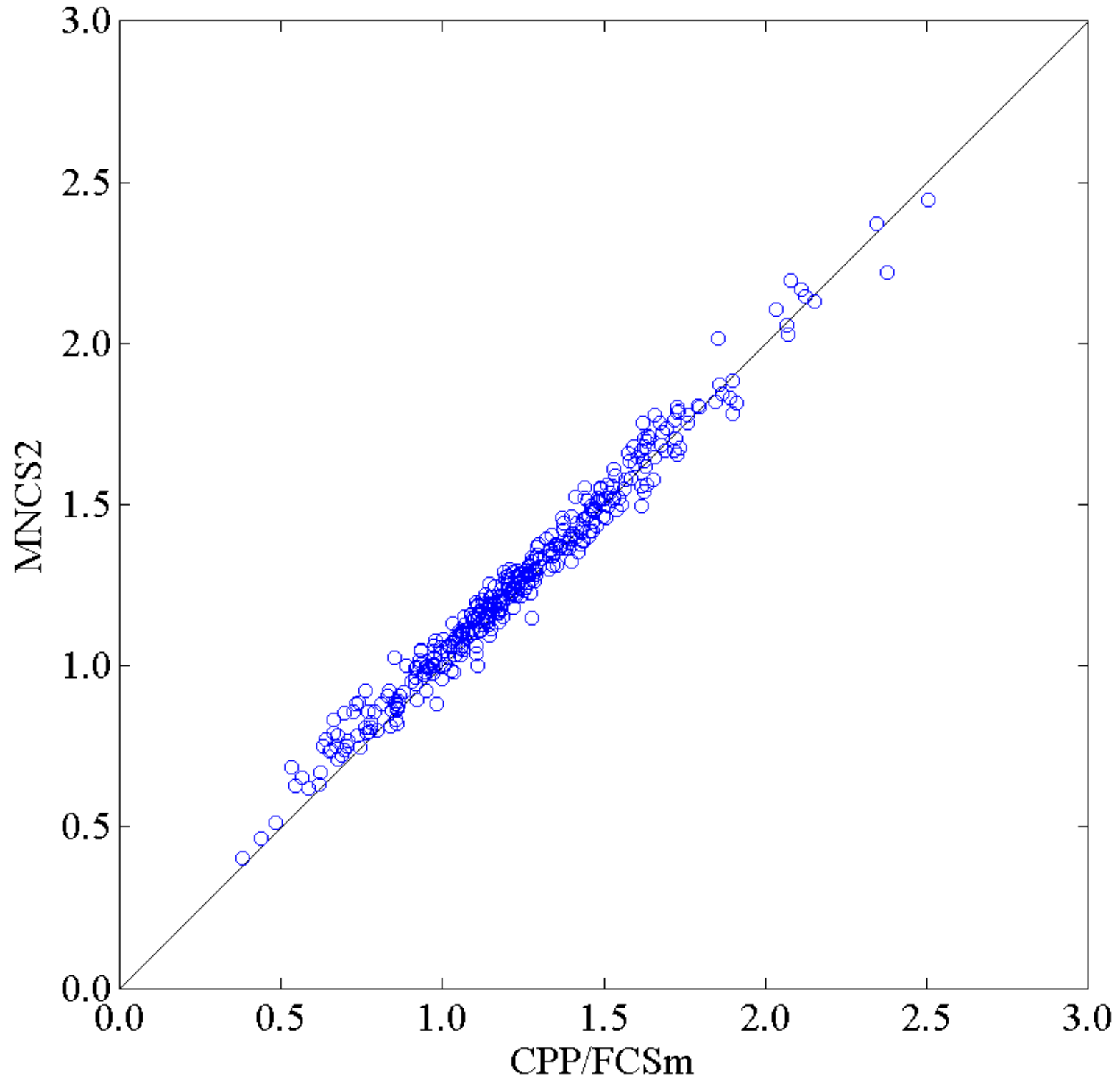

Figure 4. Relation between the CPP/FCSm indicator and the MNCS2 indicator for the research institutions data set.

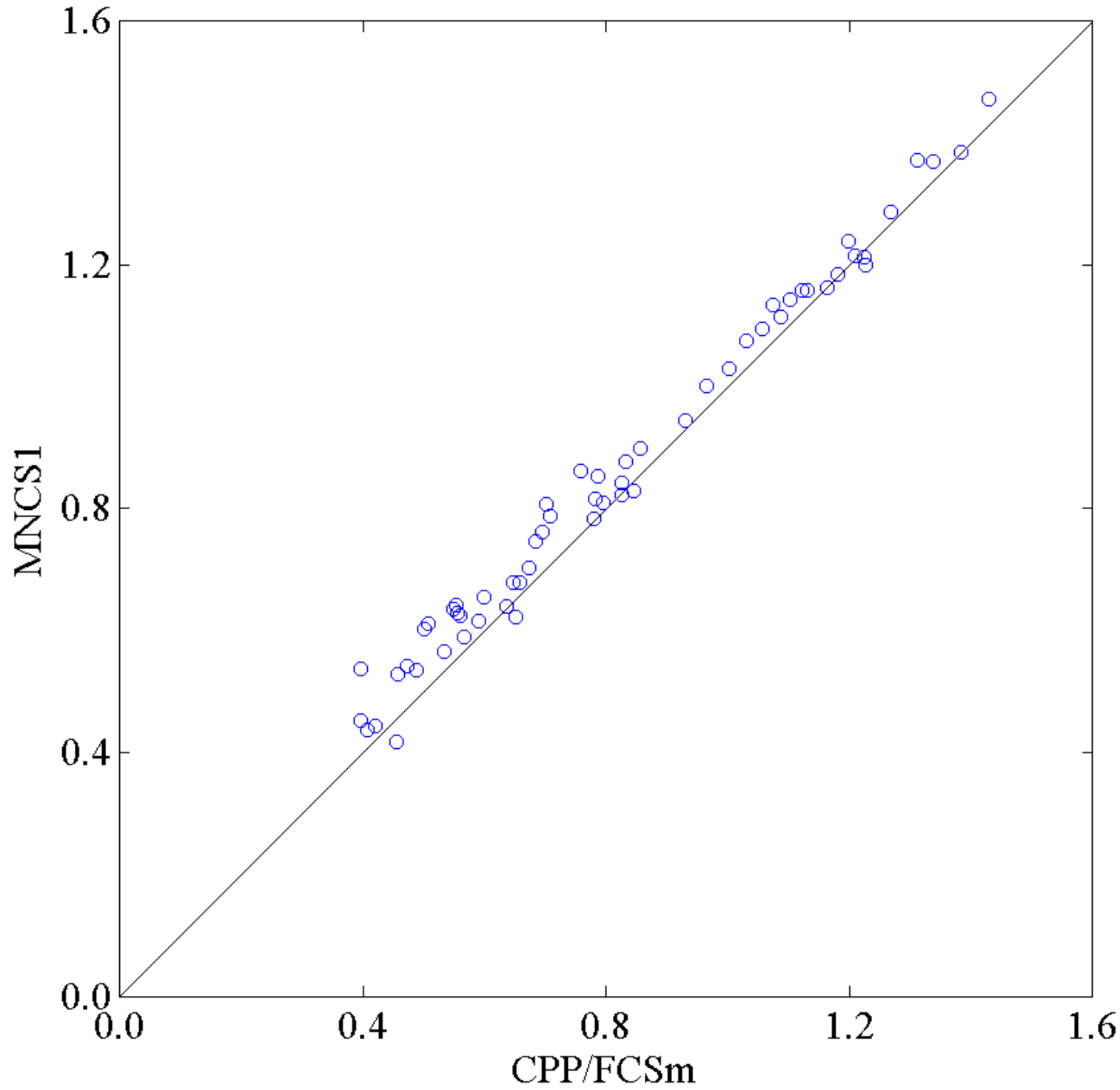

Figure 5. Relation between the CPP/FCSm indicator and the MNCS1 indicator for the countries data set.

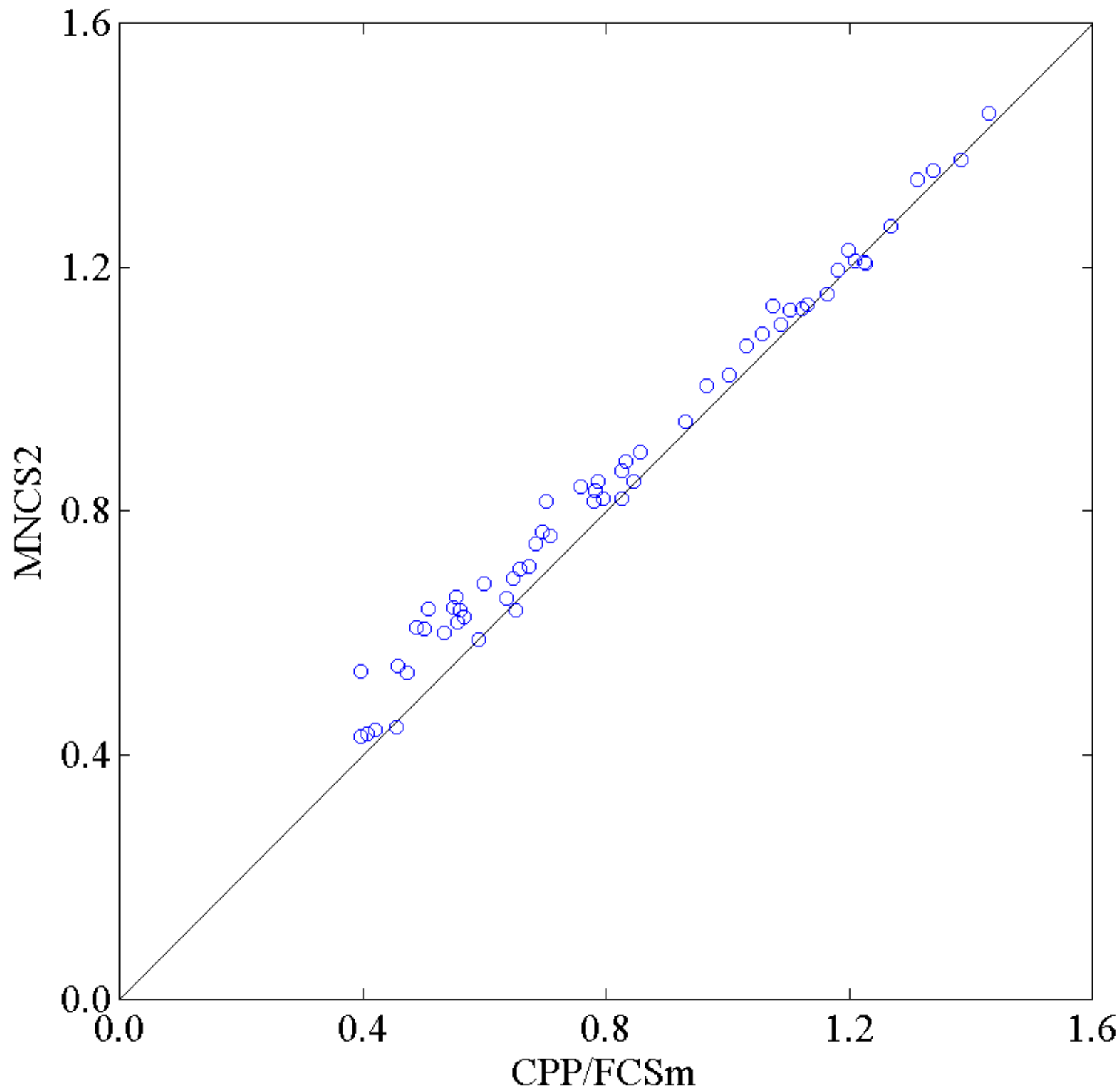

Figure 6. Relation between the CPP/FCSm indicator and the MNCS2 indicator for the countries data set.

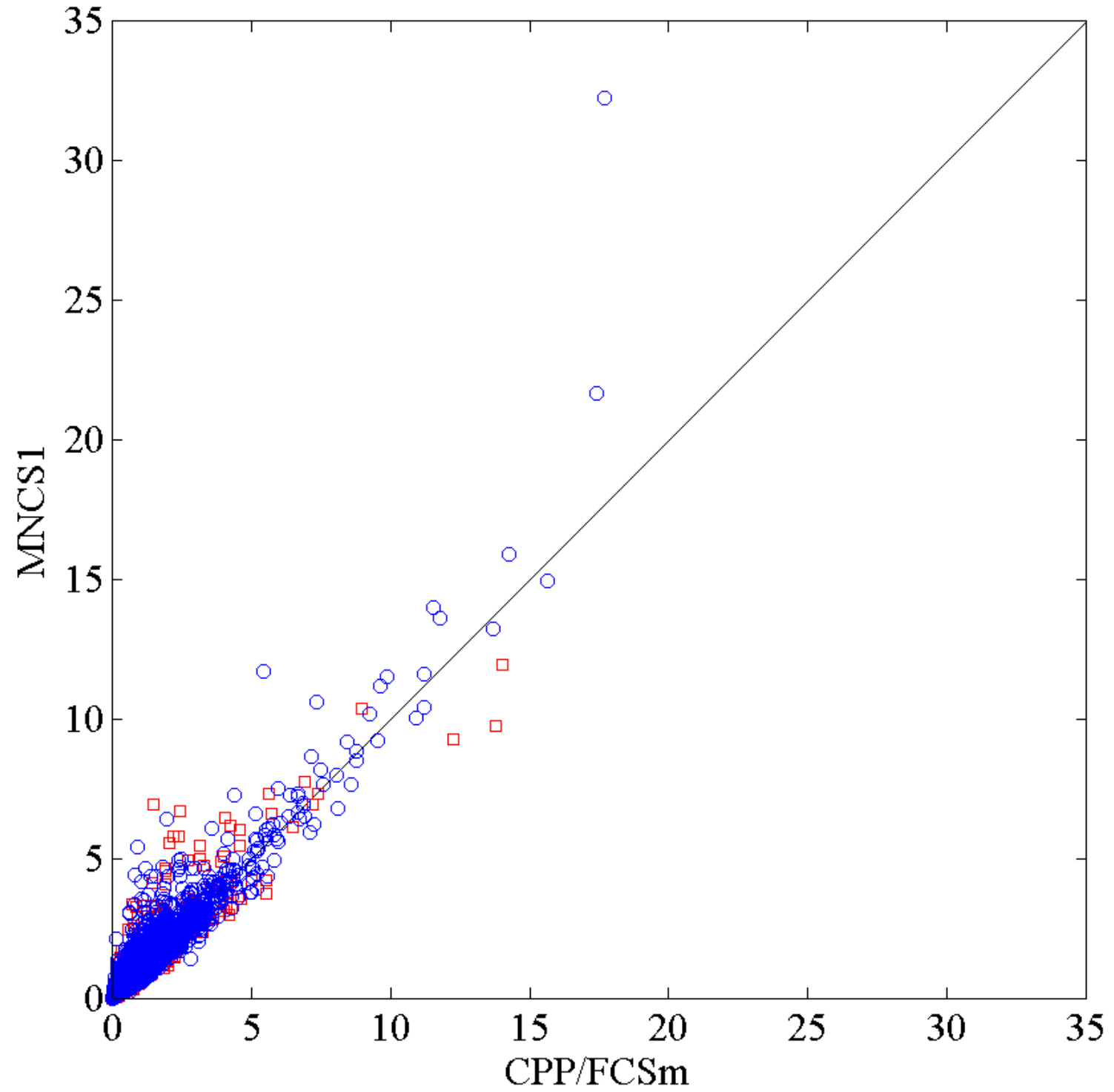

Figure 7. Relation between the CPP/FCSm indicator and the MNCS1 indicator for the journals data set.

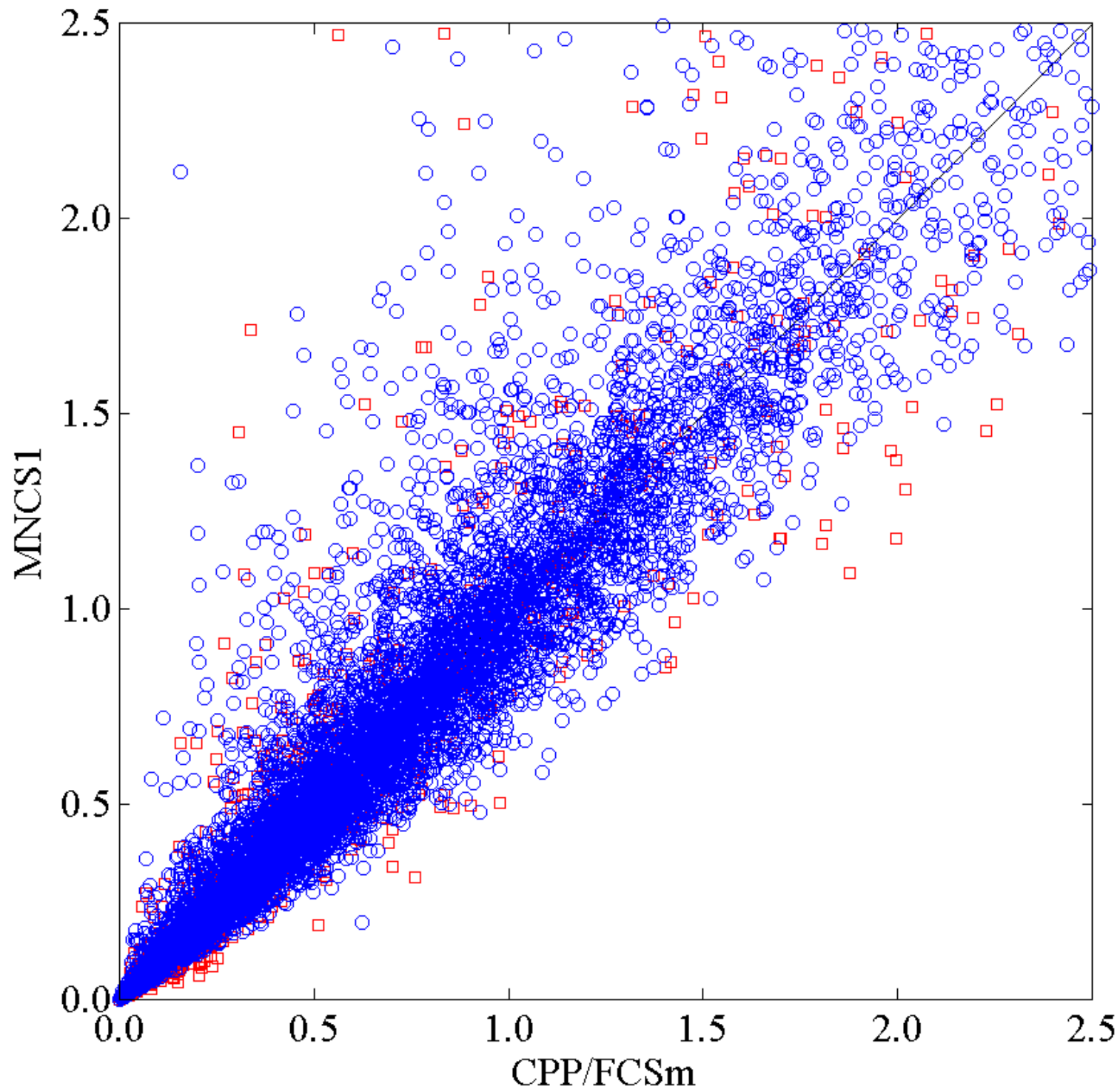

Figure 8. Relation between the CPP/FCSm indicator and the MNCS1 indicator for the journals data set. Only journals with a CPP/FCSm score and an MNCS1 score below 2.5 are shown.

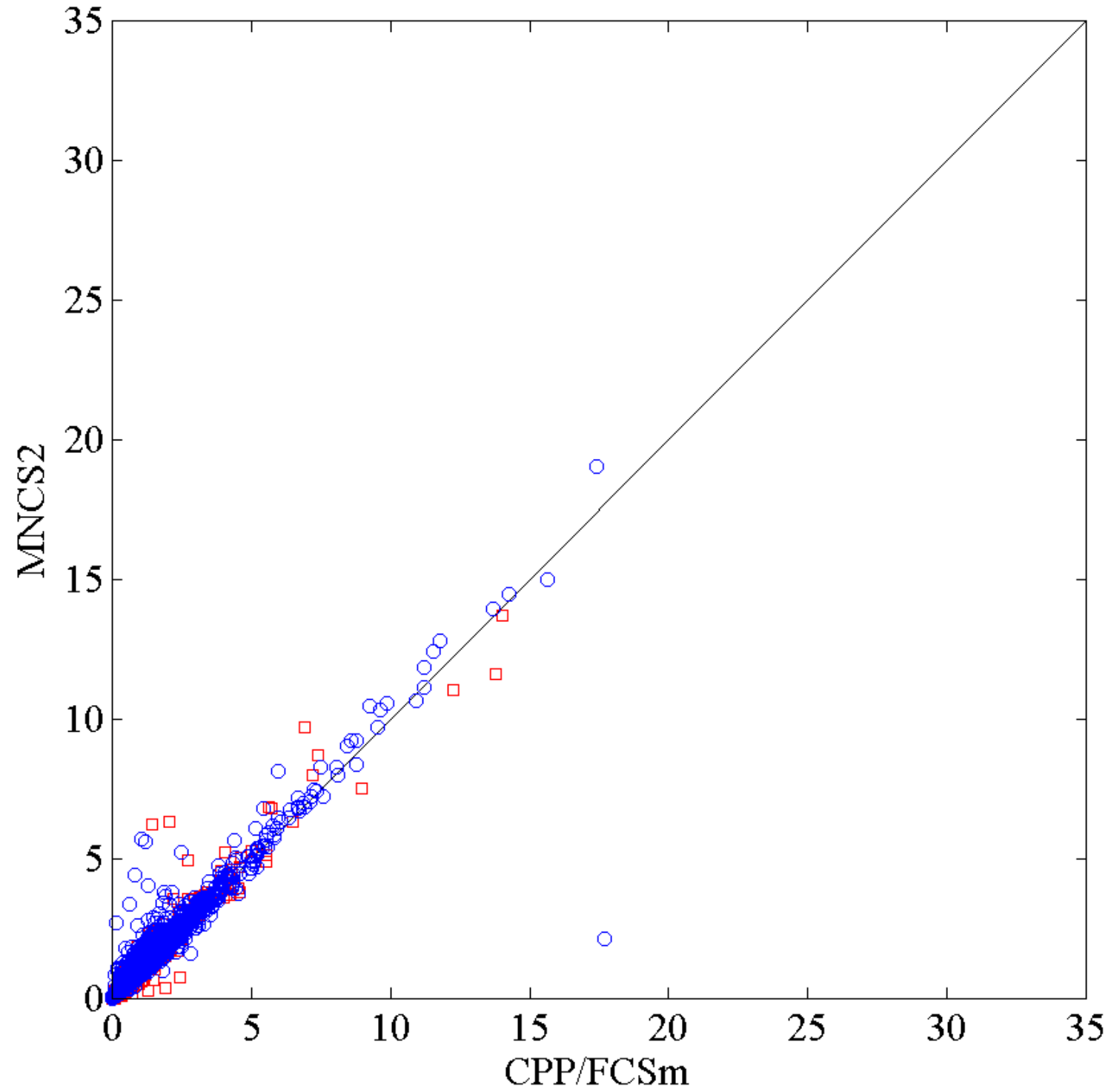

Figure 9. Relation between the CPP/FCSm indicator and the MNCS2 indicator for the journals data set.

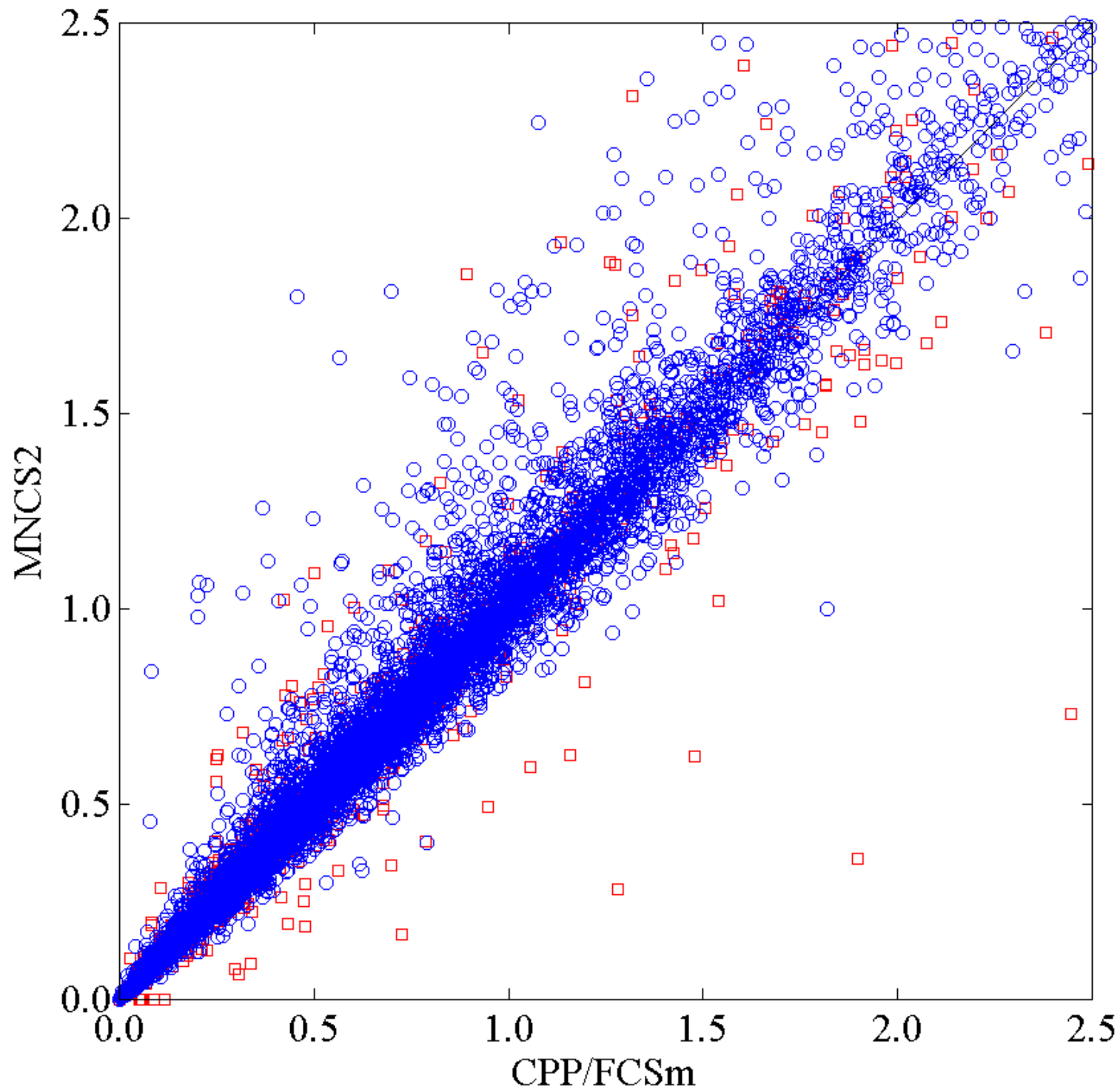

Figure 10. Relation between the CPP/FCSm indicator and the MNCS2 indicator for the journals data set. Only journals with a CPP/FCSm score and an MNCS2 score below 2.5 are shown.